# Ternary and senary representations using DNA double-crossover tiles

**Byeonghoon Kim[1,5], Soojin Jo[2,5], Junyoung Son[1,5], Junghoon Kim[1,5], Min Hyeok Kim[2],**

**Si Un Hwang[1], Sreekantha Reddy Dugasani[1], Byung-Dong Kim[3], Wing Kam Liu[4],**

**Moon Ki Kim[2,6] and Sung Ha Park[1,6]**

[1] Sungkyunkwan Advanced Institute of Nanotechnology (SAINT) and Department of Physics, Sungkyunkwan University, Suwon 440-746, Korea

[2] Sungkyunkwan Advanced Institute of Nanotechnology (SAINT) and School of Mechanical Engineering, Sungkyunkwan University, Suwon 440-746, Korea

[3] Department of Plant Science and Plant Genomics and Breeding Institute, Seoul National University, Seoul 151-921, Korea

[4] Department of Mechanical Engineering, Northwestern University, Evanston, IL 60208, USA

[5] These authors contributed equally.

[6] Address for correspondence: School of Mechanical Engineering, Sungkyunkwan University, Suwon 440-746, Korea (MKK) or Department of Physics, Sungkyunkwan University, Suwon 440-746, Korea (SHP).

E-mails: mkkim1212@skku.edu (MKK) and sunghapark@skku.edu (SHP)

**Abstract.** The information capacity of DNA double-crossover (DX) tiles was successfully increased beyond a binary representation to higher base representations. By controlling the length and the position of DNA hairpins on the DX tile, ternary and senary (base-3 and base-6) digit representations were realized and verified by atomic force microscopy (AFM). Also, normal mode analysis (NMA) was carried out to study the mechanical characteristics of each structure.

**Online supplementary data** available from stacks.iop.org/Nano/XX/XXXXXX/mmedia

DNA tiles have played a central role in the development of DNA nanotechnology. Of the many types of tiles created to date [1-8], one of the most versatile has been the double-crossover (DX) tile, having been used in a multitude of ways, from constructing one-, two-, and quasi-two-dimensional lattices [9-11] to creating space-time representations of algorithms by acting as Wang tiles [12-15]. In the case of algorithmic self-assembly, DX tiles were used as binary digits in representing data, where each of the tiles was distinguished by a presence (absence) of a hairpin representing 1-bit (0-bit) [16]. Hence, in order to increase the amount of information within a given lattice, one could increase the size of the lattice or increase the information density of each tile comprising the lattice. Yet for reasons not fully understood, lattices larger than ~1 μm in width seem difficult to grow [17]. In this work, we present a method of raising the overall information capacity of a lattice by taking the latter approach of increasing the information density of its constituent tiles. To achieve this, DX tiles with several different hairpin lengths and positions were used to represent different digits.

A DX tile consists of four strands of DNA in the form of a double duplex with two double-crossover junctions. At the 4 termini of the duplexes, single-stranded overhangs act as sticky-ends for bindings with other complementary sticky-ends (and hence tiles). Two types of DX tiles participate in the crystallization process. We follow the naming convention of these two types of DX tiles as R- and S-type tiles [18]. From the top view of the DX tile in Figure 1a, S-type tiles have two 5' sticky-ends in the upper duplex and two 3' sticky-ends in the lower duplex, and vice versa for R-type tiles. This reversed directionality between the two types of tiles forces them to bind in an alternating upside-down manner. All hairpin length and coordinate modifications were done to the S-type DX tiles, which are called the hairpin tiles. The R-type tiles are referred to as connector tiles and serve as a reference to the S-type tiles.

Hairpins were designated at 4 different sites, enumerated as 1 through 4, on a DX tile (indicated as circled numbers in Figure 1a). As can be seen in Figure 1b, three different DX tiles represent a ternary (base 3) digit, *i.e.*, DX tiles without hairpins, with 2 short hairpins (8 base pairs each), and with 2 long hairpins (16 base pairs each). In Figure 1c, six different DX tiles represent a senary (base 6) digit, *i.e.*, (0000), (0100), (1100), (0110), (1101), and (1111). The nominal scheme of the senary tiles is one in which each of the four digits represents a hairpin site, which is either 0 (without a hairpin) or 1 (with a hairpin). For instance, a (0100) tile means a hairpin exists at the second position of the hairpin modification site. Straightforward counting assumes a total of 16 different possible DX tiles (a hexidecimal representation) according to their hairpin coordination, from (0000) to (1111). Experimentally, however, several cases are indistinguishable due to their geometrical symmetry (Figure S2, see Online Supplementary Data). Within realistic experimental resolution limits of the AFM, DX tiles having 1 hairpin, *i.e.*, (1000), (0100), (0010), and (0001), are the same insofar as rotational symmetry is concerned. Hence, we use (0100) as a representative tile for DX tiles with 1 hairpin. The same argument can be applied to DX tiles with 3 hairpins, in which case we shall use (1110). In the case of DX tiles with 2 hairpins, six species of tiles exist, *i.e.*, (1100), (0011), (1010), (0110), (1001), and (0101). These tiles are classified into two types, ones having 2 hairpins on the same side normal to the DX tile or having 1 hairpin on each side of the DX tile. We use (0110) to represent both (0110) and (1001) tiles, which have 2 hairpins protruding from the same side and (1100) to represent (1100), (0011), (1010), and (0101) tiles, *i.e.*, tiles with 2 hairpins pointing in opposite directions.

All ternary and senary DX representations were successfully self-assembled under standard annealing protocols (see the materials and methods section). Each digit was expressed as a DX lattice consisting of hairpin and connector DX tiles (Figure 2). Ternary digits (Figure 2a-c) can be characterized solely by the length of the hairpins on the S-type tile. Analysis of the hairpin heights was straightforwardly carried out from the AFM data (Figure 2a-c, j) from which the height differences of the N-HP (no hairpins), S-HP (8 base pairs hairpins), and L-HP (16 base pairs hairpins) tiles could clearly be made with mean values of 1.2 nm, 2.0 nm, and 2.5 nm, respectively. For senary digits (Figure 2d-i, j), a combination of the number of hairpins and their positions serves as distinguishing attributes in representing the digits. A consistent trend appears where an increase in the number of hairpins per tile shows an increase in the height of the hairpins. A couple of factors can be attributed to this phenomenon. One is the increase in the intrinsic structural integrity of the hairpins as their numbers are increased and the other is due to the convolution effect of the AFM during scanning. Of the 6 different tiles, all but the (1100) and (0110) tile can be characterized by height measurements. The mean heights of (1100) and (0110) fall well within their respective standard deviations meaning another parameter is needed to distinguish these two types of tiles. Since the (0100) and (1100) tiles have only 1 hairpin protruding from either side of the DX duplex plane compared to the (0110), (1110), and (1111) tiles which have 2 hairpins protruding from the same plane, the width measurements of the hairpins can serve as a discriminating parameter when differentiating these tiles. The mean hairpin width for tiles with a single hairpin protrusion, $w_1$, is 9 nm. For tiles with double hairpin protrusions, the width, $w_2$, is 15 nm (Table S4, see Online Supplementary Data). This marked difference in width allows for a clear distinction of all 6 types of senary tiles.

To get a better understanding of the inherent structural characteristics of DX tiles with different hairpin lengths and coordinates, normal mode analysis (NMA) was carried out on each of the DX tiles. NMA, although auxiliary to the kinetic and thermodynamic energy scales in self-assembling DNA structures, provides useful information on the inherent vibration modes of the structure [19-21]. We worked within the framework of the mass-weighted chemical elastic network model (MWCENM) [22]. The model was created as follows. First, a network with information containing the coordinates of all the atoms was set up. This was done by reconstructing atoms of the DX tiles from crystallographic data taken from the Nucleic acid Database Bank (NDB) (see Online Supplementary Data) [23-25]. Next, representative atoms were selected by coarse-graining (Figure S3). In this model, the atoms within a cut-off distance of 8 Å of each other were treated as simple harmonic oscillators with the spring constants depending on the type of chemical interactions being taken into account, *i.e.*, van der Waals interactions, hydrogen bonds, and covalent bonds (this is what is meant by "chemical"). Here, the energy scale of these interactions was set to a ratio of 1:10:100. In addition, the masses of the atoms which surround the representative atoms, but have been left out through coarse-graining, were added to the coarse-grained atoms, forming a lumped mass. This "mass-weighted" scheme better reflects the vibrations by incorporating an inertia effect. By solving the equations of motion for this network of atoms, the inherent vibration modes of the system were obtained. The mode with the lowest energy corresponds to the most likely vibration of the system under equilibrium.

The lowest ten vibration modes of all the tiles were obtained by MWCENM-NMA (Figure S4, see Online Supplementary Data). Figure 3 shows the five lowest modes for all the tiles. The duplexes of the DX tiles are indicated in gold and the hairpins are indicated in red or blue (corresponding to the color scheme of Figure 1). The vibrations are indicated as black and empty arrows, where the same colors signify the same phase. The first mode is an out-of-plane bending (half-sine) motion for all the tiles. This is expected, since the lowest mode for cylinders is a bending motion. For all the calculated vibrations, bending (modes 1, 3, and 4) and twisting (mode 2) motions were dominant for tiles without hairpins, *i.e.*, N-HP and (0000). For tiles with hairpins, the vibrations of the hairpins are influenced by the vibrations of the DX tile and vice versa. The three main factors that affect the vibrations between the body duplexes and hairpins are the number, length, and coordination of hairpins. For tiles with short hairpins (8 base pairs), the vibration of the body did not seem to be considerably affected by the motion of the hairpins. This can be seen by comparing the modes of the (0000) and (0100) tiles. Furthermore, the apparent features of the duplex and hairpin motions of the senary digit tiles were altered by the number of hairpins. The greater the number of hairpins, the greater the deviation from the usual duplex vibrations shown by (0000) tiles. Also, the hairpin coordination, especially the symmetry of the hairpins with respect to the plane of the body, gave rise to different vibration modes of the body. For instance, for tiles with symmetric hairpin coordinations, *e.g.*, (1100), twisting modes of the body appeared much more frequently compared to the other modes. Moreover, the motions of the hairpins were also symmetric for these cases. On the other hand, for tiles with asymmetric hairpins, *e.g.*, (0110) and (1110), the vibrations of the body showed hybrid motions of twisting and bending simultaneously such as modes 3, 4, and 5 of (1110). For tiles with long hairpins, the length of the 2 hairpins above and below the DX duplex plane equals that of the body of the duplex (~32 base pairs). Hence, the amplitude of the motions of the hairpins is much greater than that of the shorter hairpins. Of particular note, a unique vibration motion appeared for mode 5 of the L-HP tile. In this mode, the body oscillated with respect to the axis penetrating the center of the two duplexes of the body.

In conclusion, we have successfully increased the information density of DX tiles by designing ternary and senary digit representations beyond the conventional binary representation of DX tiles. By controlling and varying the structural features of the hairpins, the amount of information embedded in a single DX tile was increased by a factor of $\log_2 3$ (~1.58) and $\log_2 6$ (~2.58) for ternary and senary digit tiles, respectively. This increase in information capacity allows more information to be represented in a smaller lattice as well as more diverse logic rules to be implemented. In addition, normal mode analysis was performed for all the DX tiles to analyze their inherent structural characteristics. The theoretically obtained normal modes and vibrational frequencies not only provide information on the basic vibrations of the DX tiles, but can also serve as a reference for the future work in experiments characterizing the physical properties of DNA by methods such as Raman spectroscopy.

**Materials and methods**

All oligonucleotides used in this experiment were synthesized and purified by high-performance liquid chromatography (HPLC) by Bioneer Co. Ltd (Daejeon, Korea). DNA structures were formed by mixing a stoichiometric quantity of each strand in physiological 1×TAE/$Mg^{2+}$ buffer, *i.e.*, Tris-Acetate-EDTA (40 mM Tris, 1 mM EDTA (pH 8.0)) with 12.5 mM magnesium acetate. The annealing process of the DX DNA structures started from 95 °C to room temperature for 24 hours by placing the test tube in 1.5 L of boiled water in a Styrofoam box to facilitate DNA hybridization. The final concentration of all DX DNA structures was 200 nM. AFM measurements were done by fluid tapping mode. First, 45 μL of 1×TAE/Mg$^{2+}$ buffer was deposited onto a freshly cleaved mica surface. Secondly, 5 μL of DNA solution was dropped onto the mica and this sample was placed onto the AFM stage. AFM images were obtained on a Digital Instruments Nanoscope III (Vecco, USA) with a multimode fluid cell head in tapping mode under a buffer using NP-S oxide-sharpened silicon nitride tips (Vecco, USA).

**Acknowledgement.** This work was supported by the National Research Foundation of Korea (NRF) funded by the Ministry of Science, ICT & Future Planning (MSIP) (2011-0014584) to MKK and (2012M3A7B4049801, 2012-005985) to SHP. MKK also wishes to acknowledge the support of World Class University program funded by the Ministry of Education, Science and Technology (R33-10079).

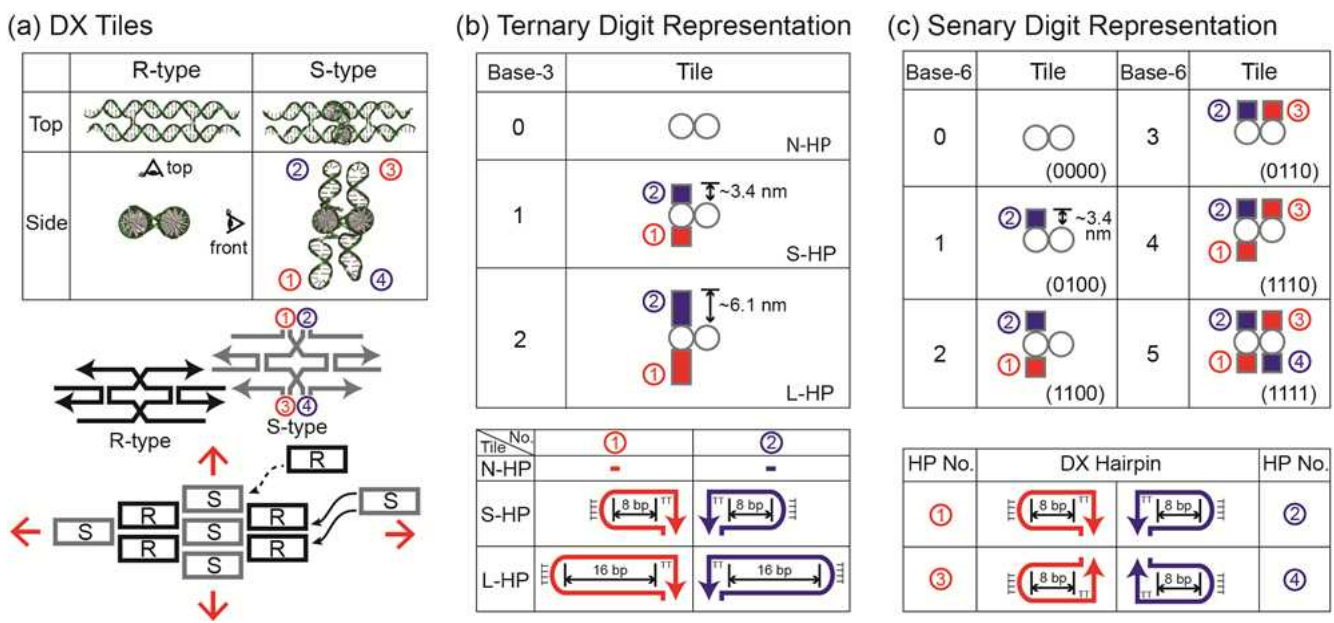

**Figure 1.** Schematic diagrams of DX tiles. The hairpin locations on S-type DX tiles are numbered from 1 to 4 with numbers of the same color indicating positions on the same strand. (a) Top and side perspectives of the two types of DX tiles used in this experiment, R-type and S-type, are shown. R- type tiles were used as connector tiles and all hairpin length and coordination modifications were done to the S-type tile. A single sticky-end binding between an R- (black) and S-type (gray) tile. The designs of the tiles are such that the strands of the two types of tiles have opposite directionality. Once nucleation occurs, crystal growth proceeds in all directions. In the crystallization process, it is energetically more favorable for a tile with two possible binding domains, such as the S-type tile shown with two solid line arrows, to associate with the crystal than it is for a tile with only one possible binding domain, such as the R-type tile shown with a single dashed arrow. (b) Ternary digits (base-3 numerals) represented as DX tiles. The three digits are expressed as S-type DX tiles with hairpins of differing lengths, *i.e.*, no hairpin (N-HP), short hairpin (S-HP, 8 base pairs), and long hairpin (L-HP, 16 base pairs). (c) Senary digits (base-6 numerals) are represented as S-type DX tiles with hairpins at different positions within the tile. There are 4 locations within a DX tile in which hairpins can be placed for a total of 16 possible DX tiles. Symmetries and experimental constraints reduce that number to 6 different DX tiles distinguishable by AFM. Each digit is identified by a sequence of 4 binary digits. A 0 (1) indicates an absence (presence) of a hairpin at the location. For example, (1100) represents a DX tile with hairpins in the ① and ② positions, (0110) represents a DX tile with hairpins in the ② and ③ positions, and so forth.

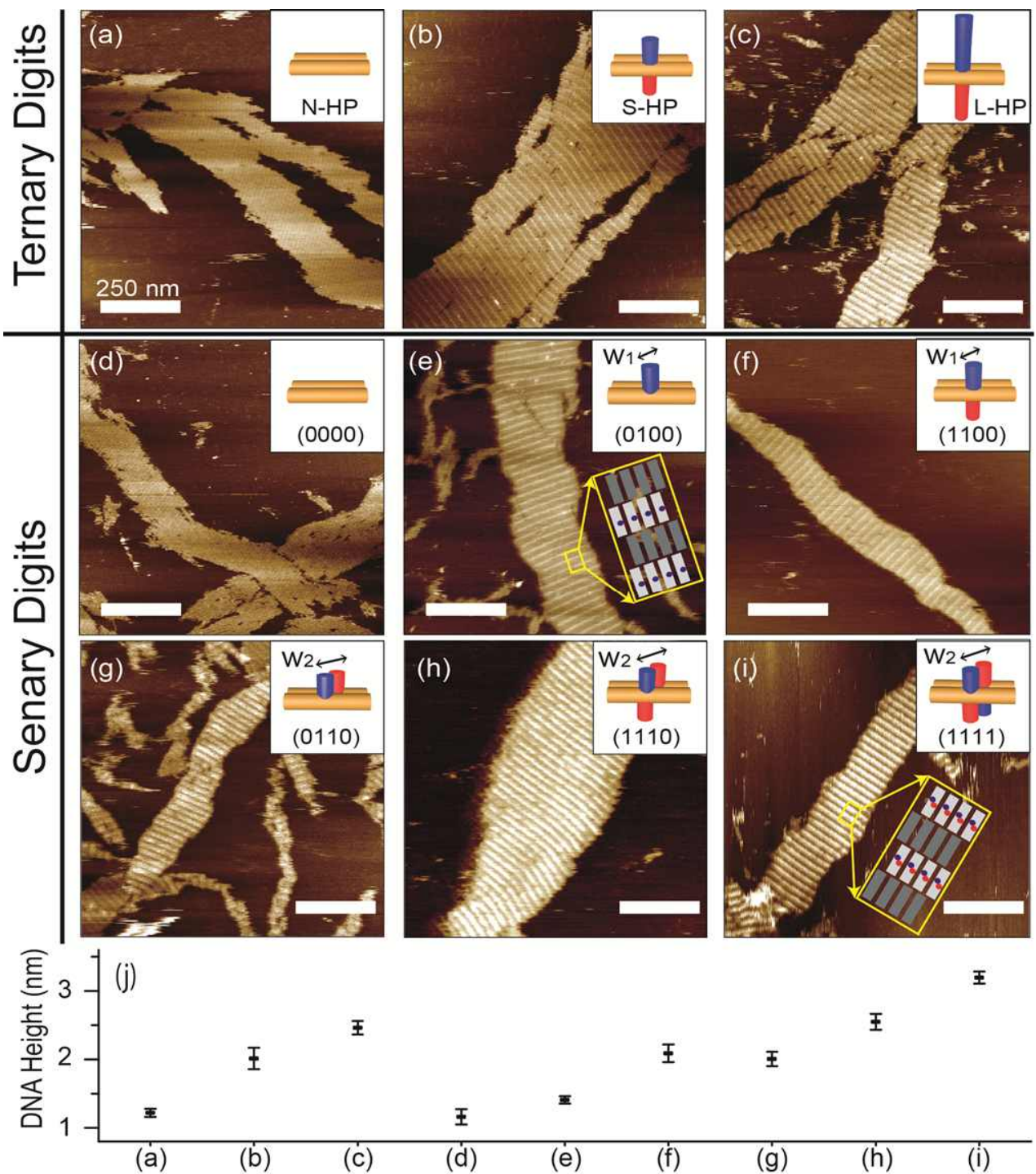

**Figure 2.** Experimental results for ternary and senary digit representations. (a-c, j) DX tiles representing ternary digits, *i.e.*, (a) N-HP, (b) S-HP, and (c) L-HP, form DX crystals which can be differentiated by the height of their hairpins. (d-i, j) Senary digits also display a similar trend. Of the 6 different tiles, all but two (*i.e.*, (f) (1100) and (g) (0110)) can be uniquely characterized by height alone. To distinguish between (1100) and (0110), hairpin width measurements through AFM data need to be taken into account. The mean width of (1100), $w_1$, is 9 nm whereas for (0110), $w_2$, it is 15 nm (see Table S4, in Online Supplementary Data). With hairpin height and width information, all 6 representations of senary digits using DX tiles can be classified. (Scale bars: 250 nm)

**Figure 3.** The five lowest modes of all the digit representations obtained by normal mode analysis (NMA). Arrows of different types (solid black and empty arrows) indicate vibrations with a phase difference of π, further depicted by the opacity of the structure. All tiles are shown from either top, side, or front viewpoints (as shown in Figure1a). The most prevalent types of vibrations for the DX tile bodies are out-of-plane sinusoidal bending motions (*e.g.*, mode 1 of all the tiles), twisting motions (*e.g.*, mode 2 of N-HP and mode 3 of S-HP), and lateral bending motions (*e.g.*, mode 4 of N-HP). The vibration modes of the hairpins show similar motions. See Figure S4 in Online Supplementary Data for full results.

# Ternary and senary representations using DNA double-crossover tiles

Byeonghoon Kim[1,5], Soojin Jo[2,5], Junyoung Son[1,5], Junghoon Kim[1,5], Min Hyeok Kim[2],

Si Un Hwang[1], Sreekantha Reddy Dugasani[1], Byung-Dong Kim[3], Wing Kam Liu[4],

Moon Ki Kim[2,6] and Sung Ha Park[1,6]

[1] Sungkyunkwan Advanced Institute of Nanotechnology (SAINT) and Department of Physics, Sungkyunkwan University, Suwon 440-746, Korea
[2] Sungkyunkwan Advanced Institute of Nanotechnology (SAINT) and School of Mechanical Engineering, Sungkyunkwan University, Suwon 440-746, Korea
[3] Department of Plant Science and Plant Genomics and Breeding Institute, Seoul National University, Seoul 151-921, Korea
[4] Department of Mechanical Engineering, Northwestern University, Evanston, IL 60208, USA
[5] These authors contributed equally.
[6] Address for correspondence: School of Mechanical Engineering, Sungkyunkwan University, Suwon 440-746, Korea (MKK) or Department of Physics, Sungkyunkwan University, Suwon 440-746, Korea (SHP); E-mails: mkkim1212@skku.edu (MKK) and sunghapark@skku.edu (SHP)

**PDB files**. The construction of the PDB files used in NMA were done as follows. First, the relative coordinates of the atoms composing each complementary base pair (A-T, T-A, G-C, and C-G) were obtained from X3DNA (X.J. Lu and W.K. Olson, *Nature Protocols*, 2008, **3**, 1213). To construct the DX body, the sequence of the DX tiles consisting of these base pairs was translated and rotated along the axis of the duplex so as to have interval spacings of 0.34 nm and a helicity of 34.3°. Hairpin duplexes were fashioned in the same manner and translated and rotated into their appropriate positions.

**Figure S1.** DNA sequence maps of all DX tiles. The names of each strand and their sticky-end sequences are indicated in red and purple. Complementary sticky-ends are denoted by a capital "S" followed by a number, *e.g.*, S1 has complementary set with S1'.

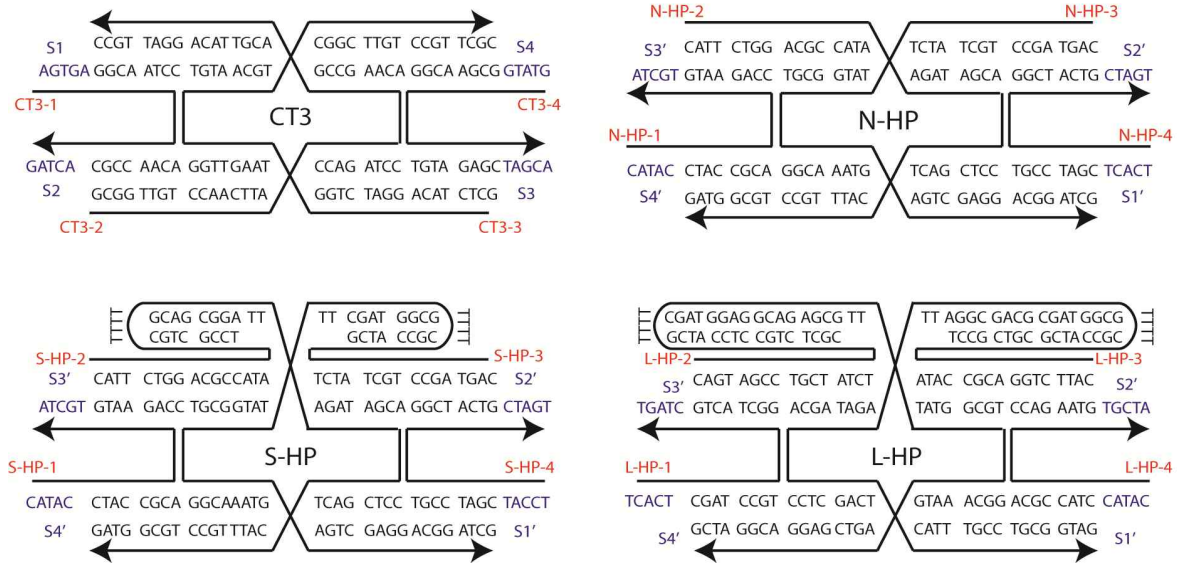

# Senary Digit Structures

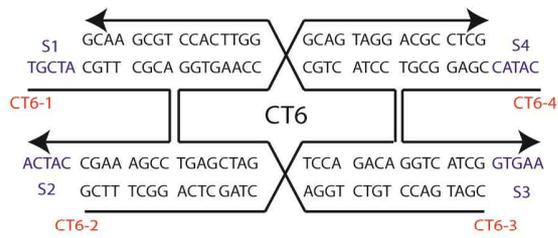
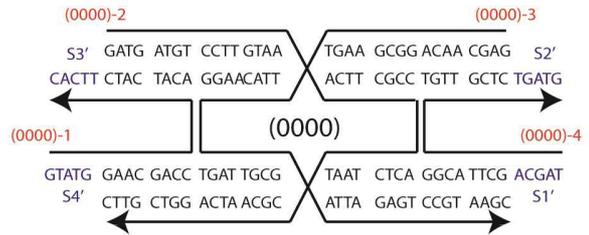
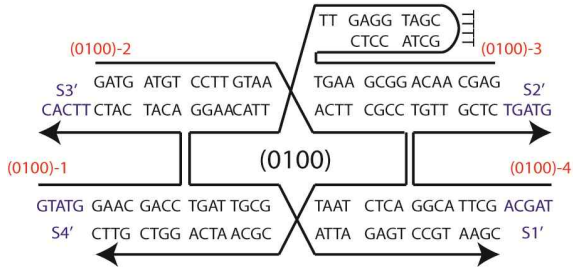
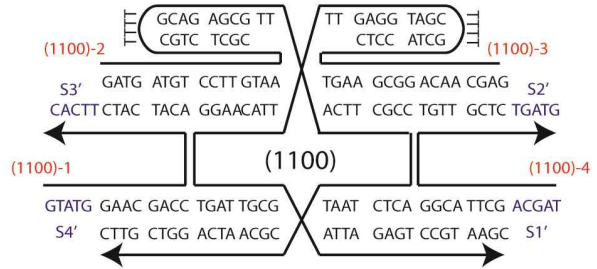
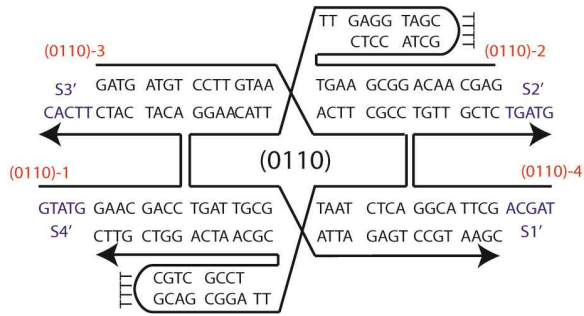
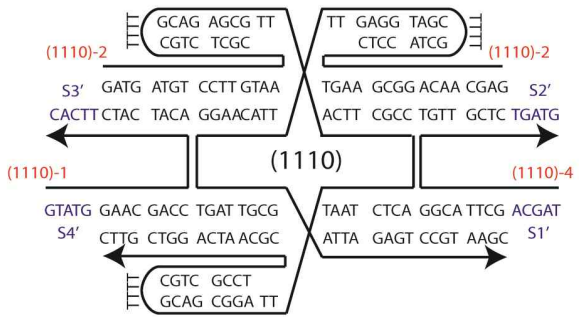
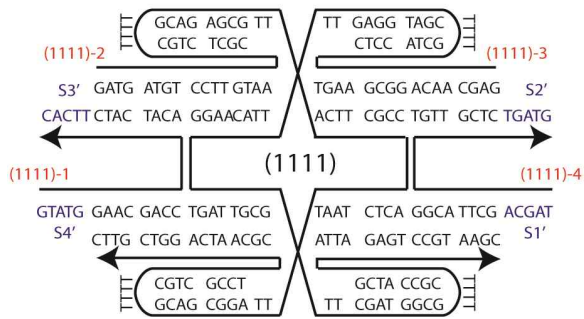

**Table S1.** Strand reference table for ternary and senary digit tiles.

| DX Tile | Strand1 | Strand2 | Strand3 | Strand4 |
|---------|---------|---------|---------|---------|
| **CT3** | CT3-1 | CT3-2 | CT3-3 | CT3-4 |
| **N-HP** | N-HP-1 | N-HP-2 | N-HP-3 | N-HP-4 |
| **S-HP** | S-HP-1 | S-HP-2 | S-HP-3 | S-HP-4 |
| **L-HP** | L-HP-1 | L-HP-2 | L-HP-3 | L-HP-4 |
| **CT6** | CT6-1 | CT6-2 | CT6-3 | CT6-4 |
| **(0000)** | (0000)-1 | (0000)-2 | (0000)-3 | (0000)-4 |
| **(0100)** | (0100)-1 | (0100)-2 | (0100)-3 | (0100)-4 |
| **(1100)** | (1100)-1 | (1100)-2 | (1100)-3 | (1100)-4 |
| **(0110)** | (0110)-1 | (0110)-2 | (0110)-3 | (0110)-4 |
| **(1110)** | (1110)-1 | (1110)-2 | (1110)-3 | (1110)-4 |
| **(1111)** | (1111)-1 | (1111)-2 | (1111)-3 | (1111)-4 |

**Table S2.** Strand information, *e.g.*, number of bases and sequences from 5' to 3' for all DX tiles.

| Strands | Number of bases | Base sequences (5' to 3') |
|---------|-----------------|---------------------------|
| CT3-1 | 26 | **AGTGA** GGCA ATCC ACAA CCGC **ACTAG** |
| CT3-2 | 48 | GCGG TTGT CCAA CTTA CCAG ATCC ACAA GCCG ACGT TACA GGAT TGCC |
| CT3-3 | 48 | GCTC TACA GGAT CTGG TAAG TTGG TGTA ACGT CGGC TTGT CCGT TCGC |
| CT3-4 | 26 | **GTATG** GCGA ACGG TGTA GAGC **TAGCA** |
| N-HP-1 | 26 | **CATAC** CTAC CGCA CCAG AATG **TGCTA** |
| N-HP-2 | 48 | CATT CTGG ACGC CATA AGAT AGCA CCTC GACT CATT GCCT GCG GTAG |
| N-HP-3 | 48 | CAGT AGCC TGCT ATCT TATG GCGT GGCA AATG AGTC GAGG ACGG ATCG |
| N-HP-4 | 26 | **TCACT** CGAT CCGT GGCT ACTG **CTAGT** |
| S-HP-1 | 26 | **CATAC** CTAC CGCA CCAG AATG **TGCTA** |
| S-HP-2 | 48 | CATT CTGG ACGC CATA TCCG CTGC TTTT GCAG CGGA TT AGAT AGCA CCTC GACT CATT TGCC TGCG GTAG |
| S-HP-3 | 48 | CAGT AGCC TGCT ATCT GCTA CCGC TTTT GCGG TAGC TT TATG GCGT GGCA AATG AGTC GAGG ACGG ATCG |
| S-HP-4 | 26 | **TCACT** CGAT CCGT GGCT ACTG **CTAGT** |
| L-HP-1 | 26 | **CATAC** CTAC CGCA CCAG AATG **TGCTA** |
| L-HP-2 | 48 | CATT CTGG ACGC CATA TCCG CTGC GCTA CCGC TTTT GCGG TAGC GCAG CGGA TT AGAT AGCA CCTC GACT CATT TGCC TGCG GTAG |
| L-HP-3 | 48 | CAGT AGCC TGCT ATCT CGCT CTGC CTCC ATCG TTTT CGAT GGAG GCAG AGCG TT TATG GCGT GGCA AATG AGTC GAGG ACGG ATCG |
| L-HP-4 | 26 | **TCACT** CGAT CCGT GGCT ACTG **CTAGT** |
| CT6-1 | 26 | **TGCTA** CGTT CGCA CCGA AAGC **CATCA** |
| CT6-2 | 48 | GCTT TCGG ACTC GATC TCCA GACA CCTA CTGC GGTT CACC TGCG AACG |
| CT6-3 | 48 | CGAT GACC TGTC TGGA GATC GAGT GGTG AACC GCAG TAGG ACGC CTCG |
| CT6-4 | 26 | **CATAC** CGAG GCGT GGTC ATCG **GTGAA** |
| (0000)-1 | 26 | **GTATG** GAAC GACC ACAT CATC **TTCAC** |
| (0000)-2 | 48 | GATG ATGT CCTT GTAA ACTT CGCC ACTC TAAT CGCA ATCA GGTC GTTC |
| (0000)-3 | 48 | GAGC AACA GGCG AAGT TTAC AAGG TGAT TGCG ATTA GAGT CCGT AAGC |
| (0000)-4 | 26 | **TAGCA** GCTT ACGG TGTT GCTC **TGATG** |
| (0100)-1 | 26 | **GTATG** GAAC GACC ACAT CATC **TTCAC** |
| (0100)-2 | 48 | GATG ATGT CCTT GTAA ACTT CGCC ACTC TAAT CGCA ATCA GGTC GTTC |
| (0100)-3 | 48 | GAGC AACA GGCG AAGT CTCC ATCG TTTT CGAT GGAG TT TTAC AAGG TGAT TGCG ATTA GAGT CCGT AAGC |
| (0100)-4 | 26 | **TAGCA** GCTT ACGG TGTT GCTC **TGATG** |
| (1100)-1 | 26 | **GTATG** GAAC GACC ACAT CATC **TTCAC** |
| (1100)-2 | 48 | GATG ATGT CCTT GTAA CGCT CTGC TTTT GCAG AGCG TT ACTT CGCC ACTC TAAT CGCA ATCA GGTC GTTC |
| (1100)-3 | 48 | GAGC AACA GGCG AAGT CTCC ATCG TTTT CGAT GGAG TT TTAC AAGG TGAT TGCG ATTA GAGT CCGT AAGC |
| (1100)-4 | 26 | **TAGCA** GCTT ACGG TGTT GCTC **TGATG** |
| (0110)-1 | 26 | **GTATG** GAAC GACC ACAT CATC **TTCAC** |
| (0110)-2 | 48 | GATG ATGT CCTT GTAA ACTT CGCC ACTC TAAT TT AGGC GACG TTTT CGTC GCCT CGCA ATCA GGTC GTTC |
| (0110)-3 | 48 | GAGC AACA GGCG AAGT CTCC ATCG TTTT CGAT GGAG TT TTAC AAGG TGAT TGCG ATTA GAGT CCGT AAGC |
| (0110)-4 | 26 | **TAGCA** GCTT ACGG TGTT GCTC **TGATG** |
| (1110)-1 | 26 | **GTATG** GAAC GACC ACAT CATC **TTCAC** |
| (1110)-2 | 48 | GATG ATGT CCTT GTAA CGCT CTGC TTTT GCAG AGCG TT ACTT CGCC ACTC TAAT TT AGGC GACG TTTT CGTC GCCT CGCA ATCA GGTC GTTC |
| (1110)-3 | 48 | GAGC AACA GGCG AAGT CTCC ATCG TTTT CGAT GGAG TT TTAC AAGG TGAT TGCG ATTA GAGT CCGT AAGC |
| (1110)-4 | 26 | **TAGCA** GCTT ACGG TGTT GCTC **TGATG** |
| (1111)-1 | 26 | **GTATG** GAAC GACC ACAT CATC **TTCAC** |
| (1111)-2 | 48 | GATG ATGT CCTT GTAA CGCT CTGC TTTT GCAG AGCG TT ACTT CGCC ACTC TAAT TT AGGC GACG TTTT CGTC GCCT CGCA ATCA GGTC GTTC |
| (1111)-3 | 48 | GAGC AACA GGCG AAGT CTCC ATCG TTTT CGAT GGAG TT TTAC AAGG TGAT TGCG TT CGAT GGCG TTTT CGCC ATCG ATTA GAGT CCGT AAGC |
| (1111)-4 | 26 | **TAGCA** GCTT ACGG TGTT GCTC **TGATG** |

**Figure S2.** Hairpin coordinations of senary DX tiles.

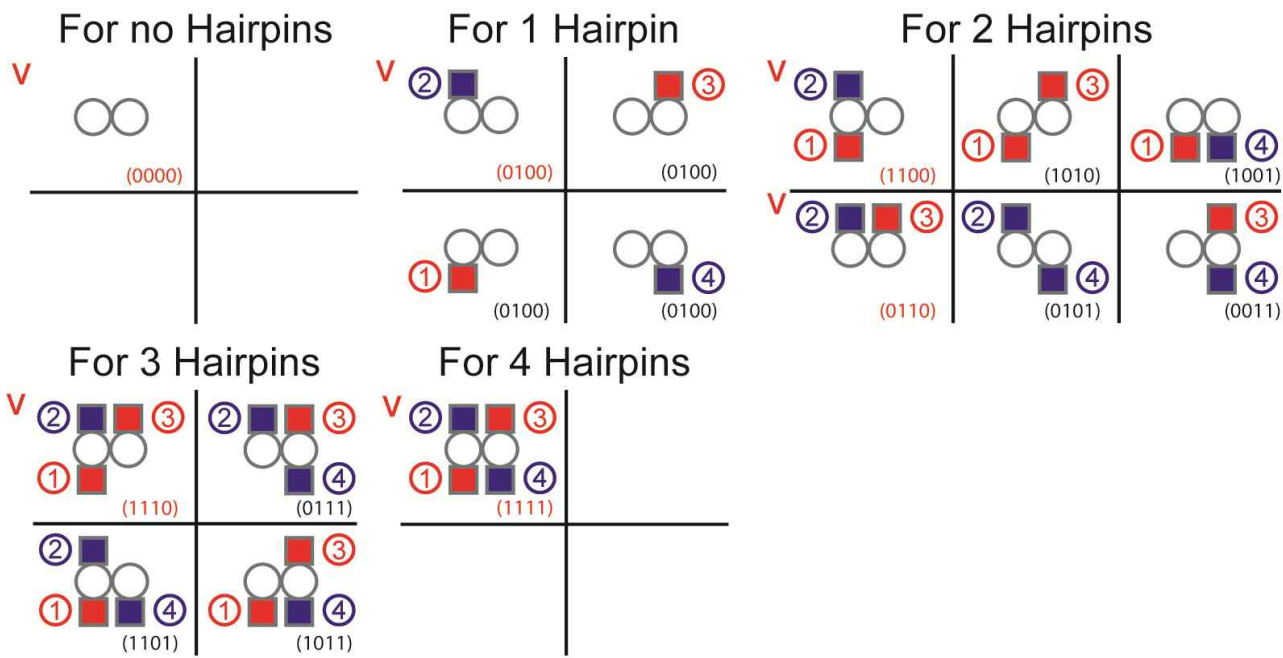

**Figure S3.** Coarse-graining scheme used in NMA. The selected atoms for each nucleotide are circled in red.

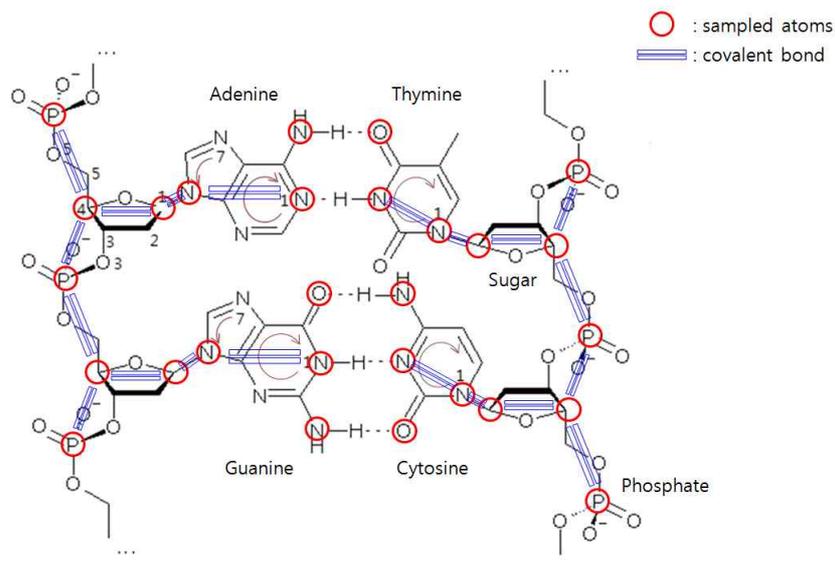

**Figure S4.** Normal mode analysis (NMA) results from mode 1 to mode 10 for all types of DX tiles. The vibrations of the DX tiles are illustrated by black and empty arrows (indicating a phase difference of π). All tiles are shown from either top, side, or front viewpoints (as shown in Figure 1a). The most frequent motions for DX tile bodies are out of plane bending motions, *e.g.*, half sine (mode 1), sine (mode 3), one-half sine (mode 5), and double sine (mode 8) of N-HP, and twisting motions, *e.g.*, modes 2 and 6 of N-HP, modes 5, 8, and 10 of S-HP. Similarly, the most common motions for hairpins were also found to be bending and twisting motions.

**Table S3.** Height analysis of all DX tiles. (units: nm)

| Counts | N-HP | S-HP | L-HP | (0000) | (0100) | (1100) | (0110) | (1110) | (1111) |
|---|---|---|---|---|---|---|---|---|---|
| 1 | 1.162 | 1.873 | 2.336 | 1.054 | 1.347 | 1.942 | 1.865 | 2.427 | 2.748 |
| 2 | 1.167 | 1.883 | 2.340 | 1.059 | 1.348 | 1.944 | 1.867 | 2.429 | 2.763 |
| 3 | 1.167 | 1.885 | 2.349 | 1.060 | 1.350 | 1.944 | 1.875 | 2.434 | 2.788 |
| 4 | 1.168 | 1.885 | 2.370 | 1.062 | 1.359 | 1.957 | 1.877 | 2.439 | 2.805 |
| 5 | 1.169 | 1.885 | 2.378 | 1.068 | 1.361 | 1.964 | 1.927 | 2.458 | 2.841 |
| 6 | 1.171 | 1.885 | 2.380 | 1.069 | 1.367 | 1.971 | 1.949 | 2.465 | 2.856 |
| 7 | 1.173 | 1.885 | 2.390 | 1.086 | 1.371 | 1.974 | 1.949 | 2.475 | 2.869 |
| 8 | 1.180 | 1.885 | 2.397 | 1.086 | 1.373 | 2.004 | 1.950 | 2.475 | 2.968 |
| 9 | 1.182 | 1.885 | 2.401 | 1.092 | 1.375 | 2.015 | 1.951 | 2.488 | 2.971 |
| 10 | 1.183 | 1.900 | 2.408 | 1.097 | 1.376 | 2.020 | 1.951 | 2.492 | 3.035 |
| 11 | 1.184 | 1.925 | 2.410 | 1.101 | 1.378 | 2.028 | 1.959 | 2.496 | 3.061 |
| 12 | 1.188 | 1.940 | 2.422 | 1.108 | 1.382 | 2.028 | 1.959 | 2.498 | 3.065 |
| 13 | 1.197 | 1.967 | 2.430 | 1.110 | 1.386 | 2.037 | 1.981 | 2.500 | 3.087 |
| 14 | 1.198 | 1.967 | 2.440 | 1.118 | 1.390 | 2.039 | 1.988 | 2.500 | 3.135 |
| 15 | 1.201 | 1.967 | 2.448 | 1.121 | 1.390 | 2.041 | 1.995 | 2.519 | 3.231 |
| 16 | 1.202 | 1.986 | 2.456 | 1.128 | 1.391 | 2.042 | 1.997 | 2.521 | 3.241 |
| 17 | 1.207 | 1.993 | 2.458 | 1.130 | 1.397 | 2.091 | 2.000 | 2.521 | 3.274 |
| 18 | 1.208 | 2.008 | 2.467 | 1.131 | 1.399 | 2.099 | 2.000 | 2.521 | 3.277 |
| 19 | 1.210 | 2.009 | 2.474 | 1.140 | 1.400 | 2.112 | 2.002 | 2.530 | 3.281 |
| 20 | 1.215 | 2.016 | 2.477 | 1.148 | 1.404 | 2.118 | 2.010 | 2.540 | 3.299 |
| 21 | 1.222 | 2.018 | 2.478 | 1.160 | 1.408 | 2.127 | 2.014 | 2.546 | 3.300 |
| 22 | 1.224 | 2.025 | 2.478 | 1.176 | 1.421 | 2.128 | 2.026 | 2.548 | 3.304 |
| 23 | 1.227 | 2.028 | 2.479 | 1.183 | 1.423 | 2.133 | 2.031 | 2.558 | 3.335 |
| 24 | 1.228 | 2.032 | 2.481 | 1.191 | 1.424 | 2.138 | 2.032 | 2.572 | 3.351 |
| 25 | 1.228 | 2.042 | 2.482 | 1.192 | 1.425 | 2.142 | 2.033 | 2.588 | 3.376 |
| 26 | 1.229 | 2.049 | 2.487 | 1.195 | 1.425 | 2.143 | 2.035 | 2.589 | 3.386 |
| 27 | 1.233 | 2.049 | 2.492 | 1.198 | 1.434 | 2.146 | 2.036 | 2.590 | 3.387 |
| 28 | 1.235 | 2.049 | 2.502 | 1.203 | 1.439 | 2.147 | 2.041 | 2.591 | 3.397 |
| 29 | 1.237 | 2.068 | 2.503 | 1.209 | 1.440 | 2.151 | 2.048 | 2.595 | 3.410 |
| 30 | 1.244 | 2.098 | 2.508 | 1.214 | 1.443 | 2.151 | 2.049 | 2.603 | 3.414 |
| 31 | 1.249 | 2.103 | 2.509 | 1.227 | 1.444 | 2.160 | 2.052 | 2.603 | 3.419 |
| 32 | 1.252 | 2.111 | 2.533 | 1.234 | 1.445 | 2.180 | 2.059 | 2.606 | 3.43 |
| 33 | 1.259 | 2.127 | 2.538 | 1.238 | 1.447 | 2.181 | 2.071 | 2.616 | 3.441 |
| 34 | 1.280 | 2.131 | 2.541 | 1.239 | 1.449 | 2.190 | 2.089 | 2.629 | 3.448 |
| 35 | 1.281 | 2.173 | 2.556 | 1.272 | 1.450 | 2.194 | 2.106 | 2.671 | 3.466 |
| 36 | 1.291 | 2.181 | 2.560 | 1.279 | 1.455 | 2.196 | 2.117 | 2.678 | 3.493 |
| 37 | 1.294 | 2.206 | 2.562 | 1.287 | 1.456 | 2.198 | 2.117 | 2.681 | 3.499 |
| 38 | 1.295 | 2.207 | 2.562 | 1.293 | 1.467 | 2.201 | 2.120 | 2.691 | 3.507 |
| 39 | 1.297 | 2.207 | 2.564 | 1.316 | 1.471 | 2.204 | 2.129 | 2.705 | 3.528 |
| 40 | 1.309 | 2.209 | 2.569 | 1.32 | 1.48 | 2.218 | 2.13 | 2.714 | 3.529 |
| Mean | 1.221 | 2.019 | 2.465 | 1.165 | 1.410 | 2.092 | 2.010 | 2.553 | 3.225 |
| Standard Deviation | 0.042 | 0.106 | 0.066 | 0.077 | 0.037 | 0.086 | 0.071 | 0.079 | 0.242 |

**Figure S5**. An example of height analysis for an S-HP crystal. All DX structures were measured from the mica surface to the top of the hairpin.

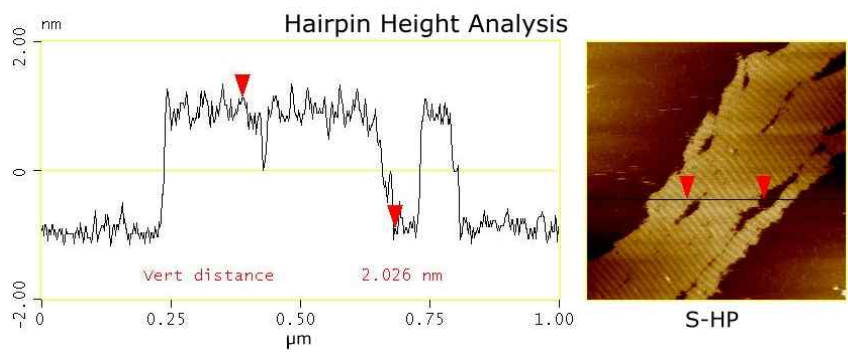

**Table S4.** Hairpin width analysis of senary digit DX tiles. (units: nm)

| Counts | (0100) | (1100) | (0110) | (1110) | (1111) |
|---|---|---|---|---|---|
| 1 | 6.809 | 6.863 | 13.725 | 12.745 | 13.725 |
| 2 | 6.809 | 6.863 | 13.725 | 12.745 | 13.725 |
| 3 | 7.782 | 6.889 | 13.725 | 13.725 | 13.725 |
| 4 | 7.874 | 6.889 | 14.153 | 13.725 | 13.725 |
| 5 | 8.366 | 7.843 | 14.153 | 13.725 | 13.725 |
| 6 | 8.755 | 7.874 | 14.706 | 14.706 | 13.725 |
| 7 | 8.755 | 7.874 | 14.706 | 14.706 | 14.271 |
| 8 | 8.755 | 8.823 | 14.706 | 14.706 | 14.271 |
| 9 | 8.858 | 8.823 | 15.163 | 14.706 | 14.271 |
| 10 | 8.858 | 8.823 | 15.163 | 14.706 | 14.706 |
| 11 | 8.858 | 8.823 | 15.163 | 14.706 | 14.706 |
| 12 | 9.350 | 8.858 | 15.163 | 14.706 | 14.706 |
| 13 | 9.350 | 8.858 | 15.686 | 15.686 | 14.706 |
| 14 | 9.350 | 8.858 | 15.686 | 15.686 | 14.706 |
| 15 | 9.727 | 9.804 | 15.686 | 15.686 | 14.706 |
| 16 | 9.727 | 9.804 | 15.686 | 15.686 | 14.706 |
| 17 | 9.727 | 9.804 | 15.686 | 15.686 | 14.763 |
| 18 | 9.727 | 9.804 | 15.686 | 15.686 | 14.763 |
| 19 | 9.727 | 9.804 | 15.686 | 15.686 | 14.763 |
| 20 | 9.727 | 9.804 | 15.686 | 15.686 | 15.686 |
| 21 | 9.727 | 9.842 | 15.686 | 15.686 | 15.686 |
| 22 | 9.842 | 9.842 | 15.686 | 16.666 | 15.686 |
| 23 | 9.842 | 9.842 | 15.686 | 16.666 | 15.686 |
| 24 | 10.700 | 10.784 | 16.174 | 16.666 | 15.686 |
| 25 | 10.700 | 10.784 | 16.174 | 16.666 | 15.686 |
| 26 | 10.700 | 10.784 | 16.174 | 17.647 | 15.686 |
| 27 | 10.700 | 10.826 | 16.174 | 17.647 | 16.666 |
| 28 | 11.673 | 10.826 | 16.174 | 17.647 | 16.666 |
| 29 | 11.673 | 10.826 | 16.174 | 18.627 | 17.647 |
| 30 | 11.673 | 10.826 | 16.666 | 18.627 | 17.647 |
| Mean | 9.471 | 9.232 | 15.350 | 15.588 | 15.027 |
| Standard Deviation | 1.243 | 1.305 | 0.814 | 1.531 | 1.098 |

**Figure S6**. An example of hairpin width analysis for a (1111) crystal. The width of all the hairpins were measured at their full width at half maximum (FWHM).

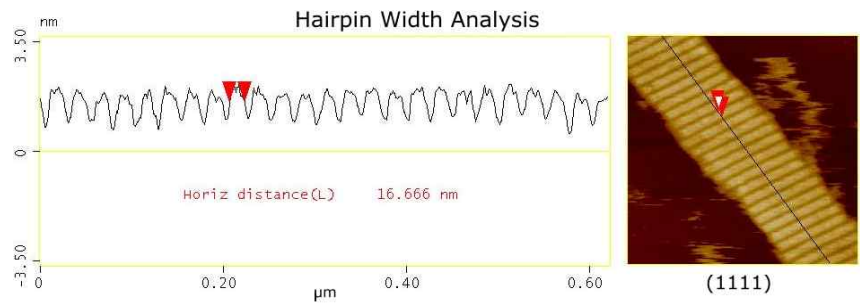